# Continuous regional trace gas source attribution using a field-deployed dual frequency comb spectrometer

**Authors:** Sean Coburn[1,*,†], Caroline B. Alden[1,2,*,†], Robert Wright[1], Kevin Cossel[3], Esther Baumann[3], Gar-Wing Truong[3], Fabrizio Giorgetta[3], Colm Sweeney[2,4], Nathan R. Newbury[3], Kuldeep Prasad[5], Ian Coddington[3], and Gregory B. Rieker[1]*

**Affiliations:**
[1]Precision Laser Diagnostics Laboratory, University of Colorado Boulder, Boulder, CO.
[2]Cooperative Institute of Research in Environmental Sciences, Boulder, CO.
[3]National Institute of Standards and Technology, Boulder, CO.
[4]National Oceanic and Atmospheric Administration, Boulder, CO.
[5]National Institute of Standards and Technology, Gaithersburg, MD.
*Correspondence and requests for materials should be addressed to: S.C. (email: coburns@colorado.edu); C.B.A. (email: aldenc@colorado.edu); or G.B.R. (email: greg.rieker@colorado.edu)
†These authors contributed equally to this work

**Abstract:** Identification and quantification of trace gas sources is a major challenge for understanding and regulating air quality and greenhouse gas emissions. Current approaches either provide continuous but localized monitoring, or quasi-instantaneous "snapshot-in-time" regional monitoring. There is a need for emissions detection that provides both continuous and regional coverage, because sources and sinks can be episodic and spatially variable. We field deploy a dual frequency comb laser spectrometer for the first time, enabling an observing system that provides continuous detection of trace gas sources over multiple-square-kilometer regions. Field tests simulating methane emissions from oil and gas production demonstrate detection and quantification of a 1.6 g min$^{-1}$ source (approximate emissions from a small pneumatic valve) from a distance of 1 km, and the ability to discern two leaks among a field of many potential sources. The technology achieves the goal of detecting, quantifying, and attributing emissions sources continuously through time, over large areas, and at emissions rates ~1000x lower than current regional approaches. It therefore provides a useful tool for monitoring and mitigating undesirable sources and closes a major information gap in the atmospheric sciences.

Emissions of greenhouse gases and pollutants pose serious risks for global climate change and human health and safety. Regional detection, quantification and attribution of trace gas sources and sinks is therefore a critical need for a variety of applications, including quantification of emissions in urban or industrial settings for monitoring, reporting, and verification; detection of small amounts of hazardous gases; verification of sub-surface sequestration efforts; and characterization of the exchange of trace gases between the atmosphere and natural or managed ecosystems. For many needs, strictly local and/or strictly time-invariant observational capabilities do not suffice for complete characterization of fluxes. For example, the "snapshots-in-time" provided by aircraft, satellite, or vehicle-mounted point sensor estimations of emissions from oil and gas operations may miss the largest fluxes, which are thought to be highly infrequent [1,2], or may misrepresent fluxes by sampling during midday, when manually triggered (operational) emissions are most frequent [3]. Similarly, regional continuous monitoring can be achieved with networks of point sensors, but the level of detail in the disaggregation of source locations and sizes must necessarily scale with the number of sensors deployed (e.g. [4]), increasing costs and complexity.

Here, we demonstrate a technology capable of continuous monitoring of trace gas fluxes, with the ability to distinguish between emissions sources at fine scales and across large areas and to infer time evolution and variability of individual sources. We present the first field deployment of dual frequency comb technology (DCS) [5,6], coupled with innovations in atmospheric inversion modeling, to enable the continuous detection, location and quantification of small trace gas sources over several square kilometer regions using a single, autonomous instrument. The system consists of the fielded dual frequency comb spectrometer, located in a centralized mobile trailer, which emits a sparse array of kilometer-scale beams strategically located throughout a region of potential emitters to sensitively measure trace gas concentrations over time (see Fig. 1). The measurements are coupled with an atmospheric transport model in a Bayesian inversion to identify sources and quantify the emission rate over time at each source location with 2-minute resolution. The laser beam is invisible and eye-safe, and the system can operate continuously day and night except during periods of total optical occlusion (e.g., heavy precipitation). Trace gas sources do not need to be imaged directly, which is important for cases in which line-of-sight from the laser to the source location is blocked by terrain or vegetation. Rather, the sensitivity of the spectrometer enables a sparse beam array that only must intersect the plumes downwind of the sources.

The frequency comb laser is a Nobel prize-winning technology [7,8] that has significantly impacted the field of molecular spectroscopy [9–12]. The femtosecond pulsed output of a mode-locked frequency comb laser is composed of 100,000+ perfectly spaced, discrete wavelength elements or "comb teeth", that act as a parallel set of continuous wave lasers with known frequencies. Dual frequency comb spectroscopy uses two combs with slightly different tooth spacing, mixed on a photodiode after transmission through the atmosphere, to extract high resolution absorption information [11]. The result is an unprecedented combination of spectral bandwidth (>100 nm), resolution (<$2 \times 10^{-3}$ nm) and signal-to-noise ratio (> $2 \times 10^4$), providing precise and accurate absorption spectra over long atmospheric paths [13,14]. The spectra are fit with a laboratory-validated absorption model to simultaneously retrieve the atmospheric concentration of all trace gases that absorb within the bandwidth. The laboratory model is the only required 'calibration', and cross validations between DCS instruments show a long-term agreement of 0.35% in $CH_4$ concentration [14], allowing for sensing networks that don't require periodic calibration. The current spectrometer is capable of detecting a range of near-infrared absorbing molecules such as $CH_4$, $H_2O$, $CO_2$, and isotopologues and could detect $O_2$, $SO_2$, $NH_3$, and CO with modifications. More complex comb spectrometers operating in the IR will expand the list of detectable molecules in the future [9].

Time-resolved location and quantification of sources with the sparse array of line-of-sight, integrated open-path measurements provided by the DCS also requires new inversion techniques. For this, we implement an inversion that identifies sources and quantifies emissions at multiple possible source locations, given a time series of observations and related covariance, a transport model to relate the sources and open-path measurements, and estimates of temporal and spatial emission and background covariance [15]. The inversion uses spectrometer measurements as the prior estimate for background concentrations, thereby removing potentially confounding signals from nearby emissions and obviating the need for additional sensors to constrain background conditions (see below and *Methods*).

**Results**

In the initial deployment described here, we choose the important case of methane emission detection and quantification from oil and gas operations to demonstrate the capability of the system. To this end, controlled methane sources are dispersed across a field site to simulate emissions from natural gas production sites. The fielded DCS is located at the Table Mountain Field Site, ~10 km north of Boulder, Colorado (Fig. 2). A trailer houses the DCS, however the volume of the DCS and supporting equipment is 0.6 x 0.9 x 0.7 m and thus amenable to smaller platforms. The launch/receive optics and pointing gimbal are mounted on the trailer roof or an adjacent tower and have been subjected to four seasons of weather over a 12-month operational period including drastic temperature variations (~18 ºC daily), significant wind loading (>30 m s$^{-1}$) and precipitation (rain, snow). Retroreflectors are placed at distances of up to 1.1 km from the spectrometer. Targeted sequentially, each retroreflector reflects laser light back to the photodetector co-located with the launch optics. The retroreflectors are placed among the potential sources (lateral offset between source and beam path is 15-60 m) for measurement of upwind and downwind integrated trace gas concentrations along sets of laser beams, enabling the estimation of background concentrations for each potential emission site and for each time step. This configuration holds potential for identification of even very small sources in regions with a high density of oil and gas operations, where ambient concentrations of methane can have high spatial and rapid temporal variability.

First, we demonstrate the identification and quantification of a very small, variable-rate emission at a distance of 1 km (Fig. 3). Atmospheric measurements begin at 09:00 local time, and continue until 07:00 the following day. At 14:05, the controlled release of 7.7 g min$^{-1}$ begins. At 18:00 the rate changes to 4.6 g min$^{-1}$, at 22:00 the rate drops again to 3.1 g min$^{-1}$, and at 00:00 drops to 1.6 g min$^{-1}$, before stopping completely at 04:55 (Fig. 3). Atmospheric CH$_4$ measurements downwind of the leak show clear enhancements when the controlled release begins. The inversion successfully predicts that no leak is present before this time (the posterior flux is within 1-σ of zero). The posterior emission estimate becomes significantly greater than zero within minutes of the true leak start, demonstrating that the system can rapidly identify the onset of emissions, a particularly important feature for intermittent sources. The posterior emission estimate remains significant for the entire leak duration, becoming indistinguishable from zero only when the controlled release is shut off at 04:55 the next day. The posterior emission rate is variable, particularly during periods of low wind speed and shifting wind directions, such as occurred between 16:00 and 20:00 (see Fig. 3 and S1), and at night, when parameterization of atmospheric stability is difficult. Over the measurement period, the root-mean squared deviation between the measured and true leak rate is 2.9 g min$^{-1}$. For comparison, this value is three times smaller than the lowest mean emissions from functioning pneumatic controllers on a well site [1]. During the period identified by the inversion as having non-zero emissions, the overall average posterior emission rate is 5.2 ± 1.6 g min$^{-1}$, which is within 1-σ of the true average emission rate of 4.9 g min$^{-1}$ (Fig. 3).

A second set of field tests assesses the ability of the observing system to locate and quantify simultaneous emissions from multiple sources. To simulate an accurate representation of the density of oil and gas production in the United States, the inversion is given prior knowledge of the spatial distribution of five well sites similar to a randomly selected section of the nearby Denver-Julesburg oil and gas basin. Of five well sites, controlled methane release points are positioned at two (Fig. 4). Eight retroreflectors create an array of beams interspersed among the sites. Measurements begin at 09:00, and controlled releases begin at both emission points at 11:30 with equal rates of 3.1 g min$^{-1}$, increasing to 3.7 g min$^{-1}$ at 13:10. Atmospheric measurements continue until both controlled releases are turned off at 18:00. The inversion identifies emissions at both sites beginning at the correct time (Fig. 4). The root mean squared deviation between the estimated and true leak strength is below 1.2 g min$^{-1}$. Equally important, the inversion also correctly identifies the three non-leaking well sites as having emissions consistent with zero.

**Discussion**
The production, transport and storage of natural gas from the more than one million active wells in the US results in both intentional and unintentional emissions of 6-12 million metric tons of $CH_4$ to the atmosphere annually [16,17]. These emissions represent lost revenue, pose risks to public safety, accelerate climate change, and, through natural gas co-emissions, lead to decreased air quality [18]. The economics of leak mitigation is complicated by the wide spatial distribution and time variability of potential leaks, making the task of locating leaks with traditional optical gas imaging and handheld sensing technologies labor intensive, costly, and unreliable [19]. Existing methane sensing technologies offer high spatial but low temporal coverage or vice versa [20]. Satellite and aircraft mass balance approaches cover large regions but at coarse spatial and temporal resolution. Additionally, these methods are effective only under a subset of atmospheric conditions (e.g., clear sky) and are limited to identification of leaks greater than 1000 – 10000 g min$^{-1}$ [21–24]. Sensors mounted on vehicles require operators and offer snapshots in time [25–28]. Fixed, continuous ground-based sensors do not acquire sufficient information to locate specific sources from more than a few hundreds of meters [19], and are currently too expensive for adequate monitoring of oil and gas operations. The dual-comb spectrometer and atmospheric inversion approach demonstrated here offers the ability to continuously and autonomously monitor many potential sources across multiple square kilometer regions with emission rates down to 1.6 g min$^{-1}$. Achieving this level of sensitivity means that the system is capable of detecting all sources relevant to oil and gas infrastructure, from so-called "super-emitters", or large point sources that account for a substantial portion of annual renegade emissions, to small sources <1 ton yr$^{-1}$ (e.g., faulty pneumatic valves). Additionally, the ability to support continuous monitoring increases the chances of detecting large (and small) episodic emission sources, for which there is currently little to no data describing the frequency of occurrence. Thus, in regions of dense oil and gas operations, this approach could lead to drastically reduced monitoring costs, enabling economically viable leak mitigation.

Future applications of the observation and inversion framework described here range from detection and quantification of trace gas sources over large urban and rural regions, to sensitive early-warning systems for the presence of small amounts of airborne chemical constituents, to confirmation and monitoring of underground storage or sequestration of gaseous materials. The system bridges a critical gap in existing trace gas monitoring capabilities by providing highly sensitive, time-varying, continuous, regional-scale coverage.

## Methods
### Dual frequency comb spectrometer

The dual comb spectrometer employed here consists of two fully self-referenced erbium fiber frequency combs operating at 200 MHz [5]. Phase coherence between the two frequency combs is achieved by locking the carrier offset frequency ($f_{ceo}$) using $f$-to-$2f$ locking, and phase-locking an individual tooth from each comb to a common 1 kHz linewidth continuous wave (CW) commercial diode laser [6]. Light from each comb is spectrally shifted to the 1.65 micron region using highly nonlinear fiber, combined with the other comb, and spectrally filtered using a custom fiberized interference filter resulting in a relatively flat spectrum spanning 1.62-1.69 microns, the optimal near-infrared region for measurement of atmospheric $CH_4$ and water vapor over long paths with high precision. The filtered light is then transmitted via 20 m of single mode fiber (SMF) to the telescope transceiver which transmits to and receives light from the retroreflectors.

A single 100-MHz-bandwidth InGaAs photodetector mounted on the telescope transceiver is used for detection of the dual-comb interference signal. The detector signal is transmitted to the data collection system inside the mobile laboratory. A bias tee separates the RF and DC components of the signal. The DC portion is used to monitor the power reaching the detector. The RF portion is passed to the data collection system and digitized at 14 bits and 200 MHz (clocked at the repetition rate of one of the combs). Prior to digitizing, the dual comb signal is amplified and attenuated in order to optimize linearity of the detection system [14]. The digitizer is controlled by custom acquisition code that allows for real-time averaging of individual interferograms as well as phase correction and additional averaging of phase-corrected interferograms in order to reduce the final data burden. For these tests, individual interferograms are recorded at ~630 Hz and averaged for 128 s with phase corrections applied to the interferograms every 150 ms. An example transmission spectrum from the DCS instrument, covering the fitting region used for this study, is shown in SI Fig. 3 panel (a).

Extraction of each species concentration with high stability and low interference from overlapping molecules and etalon interference is achieved through simultaneous multi-species fits to the broadband, high resolution spectra (SI Fig. 3, panel (b)) with well-vetted temperature and pressure-dependent models of the absorption fingerprints of each molecule [13]. The combination of the dual-comb instrument and fitting approach produces results that are undistorted by atmospheric turbulence, free from instrument-specific lineshapes, robust against species interference, and requires no periodic calibration (the absorption model serves as the permanent calibration for all instruments) [13,14]. Thus, the instruments can be easily networked and the measurements linked (through the absorption model) to international standards.

### Return power
Due to various contributing factors such as turbulence and small telescope alignment drifts, there is a large range in return power. The total range is ~0-300 $\mu$W and the average is between 100-150 $\mu$W measured at the detector. The return power to achieve optimal signal-to-noise ratio (SNR) is ≥200 $\mu$W, but measurements are feasible down to ~15 $\mu$W [14].

### Spectral fitting

The following steps are taken to determine concentrations of the measured species from the collected interferograms: 1) the FFT is taken of the averaged interferogram; 2) a piece-wise polynomial is fit to the baseline of the spectrum (this process also takes into account any absorption lines of species within the fitting region); 3) the polynomial fit is removed from the spectrum; 4) an absorption model is generated (using HITRAN 2008) for the region of interest and concentrations of absorbing species are fit simultaneously with a low order polynomial [13]. Supplemental Fig. 3 panel (a) contains the raw transmission spectrum (showing only the fitting region) and panel (b) contains the model fit to the absorbance data for an example 128 s spectrum collected during the 1 km beam path tests. The raw spectrum shown in panel (a) demonstrates the capability of the DCS system to achieve high signal-to-noise measurements (peak-to-peak noise values of $3 \times 10^{-3}$ with a standard deviation of $5 \times 10^{-4}$ in absorbance units). Supplemental Fig. 3c also shows the Allan deviation plots for $CH_4$ from first open path DCS measurements [13] and the field-deployed system under windy, well-mixed conditions. Both curves follow the expected √(averaging time) decrease at short averaging times, but the fielded DCS delivers improved precision values for a given averaging time (i.e., 3 ppb vs 8 ppb in 30 s), primarily because of improved transceiver optics. Both the instruments reach a similar minimum precision value of ~2 ppb, which is the approximate stability of the atmospheric methane concentration over the seconds-to-minutes timeframe under well-mixed conditions.

**Transceiver and retroreflector placement**
During the 2-leak tests, the telescope transceiver was placed on a small platform on top of the mobile laboratory, which resulted in a total height of ~4.3 m. However, this placement was found to be suboptimal for beam path distances >600 m due to increased wind-loading on the mobile laboratory (the large sail area of the trailer resulted in increased side-to-side movement under wind speeds greater than ~4 m s$^{-1}$). Based on these observations, during the 1 km beam path test the telescope transceiver was placed on a stand-alone tower located 1-2 m northwest of the mobile laboratory. The total height of the telescope transceiver once located on the tower was ~3.7 m.

The retroreflectors used throughout this study consist of a solid aluminum hollow corner cube form coated with unprotected gold on the cube faces, known as replicated hollow corner cube retroreflectors. These optics have a clear aperture of 63.5 mm and a stated return beam accuracy of 30 arc sec. For both testing scenarios presented here, the retroreflectors are placed on standard camera tripods and had a total height between 1-1.5 m.

**Controlled methane release**
The methane leak tests in this study use standard compressed methane cylinders with the output regulated by an Alicat mass flow controller (MFC) (model MC-20SLPM-D). For the 1 km beam path tests, the cylinders and MFC are co-located and only a short line of standard ¼" outer diameter Teflon tubing is run to the leak location (<3 m of tubing). However, for the multi-source study the cylinders are placed closer to the mobile laboratory and Teflon tubing is run to the MFC, which is located at the mid-point between the two leak locations (each individual line was ~200 m). A splitter after the MFC directs the gas flow to the two individual leaks and the total flow is set to double the desired flow rate at each location. Supplemental Figs. 1 and 2 show time series of the retrieved species concentration and temperature for the 1 km and 2-leak tests, respectively.

**Inversion Framework and Parameterization**

We use a Bayesian inversion to solve for fluxes. Following [15], the standard formulation for the mass emission rate estimate, or flux estimate, $\hat{\mathbf{s}}$, is:

$$\hat{\mathbf{s}} = \mathbf{s}_p + \mathbf{QH}^T(\mathbf{HQH}^T + \mathbf{R})^{-1}(\mathbf{z} - \mathbf{Hs}_p). \tag{1}$$

The $m \times 1$ posterior flux vector is $\hat{\mathbf{s}}$. $\mathbf{s}_p$ is the $m \times 1$ state vector of prior source estimates, $\mathbf{z}$ is the $n \times 1$ vector of observations, $\mathbf{R}$ is the $n \times n$ matrix of observation covariance, $\mathbf{Q}$ is the $m \times m$ matrix of prior flux covariance, and $\mathbf{H}$ is the $n \times m$ matrix of source-receptor functions. The dimension $n$ is equal to the number of observations. The dimension $m$ is equal to the number of mass emission rates to be estimated, equal to the number of time steps evaluated multiplied by the number of potential source locations to be monitored. The Gaussian plume model (see below) offers a steady-state solution to atmospheric transport, such that the number of time steps of flux estimation is equal to the number of atmospheric observations, $n$. Assumptions of steady state atmospheric transport, based on mean meteorological conditions during a 2-minute measurement window, are an appropriate choice for the observing system described here because the travel time (approximated using mean wind speed) from a given source location to its assigned downwind beam is shorter than measurement averaging times.

**Transport operator, H**
The transport operator can be generated with any simulation of atmospheric motion. The transport operator is defined as the integrated relationship between each source and each receptor, or beam, which is discretized into ~1 m sections for calculation of $\mathbf{H}$. Here, we use a Gaussian plume model with an adaptation of Turner's method [29] for characterization of atmospheric stability (using the Pasquill-Gifford classes) driven by meteorological measurements (see Meteorological Data section above and Figs. S2 and S3) from a 3D sonic anemometer. The method is also amenable to the use of transport fields generated by high-resolution computational fluid dynamics simulations, which come at higher computational cost.

**Observation vector, z**
The vector of observations, $\mathbf{z}$, is populated with concentrations measured on beams downwind of each source location. The vector of background concentration estimates (the portion of $\mathbf{s}_p$ devoted to the background) is populated with concentrations measured on beams upwind of each source location. Upwind and downwind measurements are paired based on tolerances that the wind direction does not shift by more than 180° and that no more than 12 minutes elapse between measurements.

**Model-data mismatch matrix, R**
Parameterization of the model data mismatch term, $\mathbf{R}$, accounts for and describes measurement and transport uncertainties and their temporal covariance. The mean value of the diagonal elements of $\mathbf{R}$ is 11.2 ppb$^2$ for the single, 1 km path study. The mean value of the diagonal elements of $\mathbf{R}$ is 21.9 ppb$^2$ for the multiple-leak study. The diagonal elements of $\mathbf{R}$ vary measurement-to-measurement (see Table S1) because of changes in the measurement uncertainty term, which is the goodness-of-fit of the fits to absorption line shapes for each measurement. The increase in the model data mismatch for the multi-leak case is due to differences in atmospheric conditions (turbulence) during the measurements as well as to smaller impacts of transport uncertainty on longer beam paths, where the signal from a discrete plume is smaller on a longer beam path than on a shorter beam path. **R** includes measurement uncertainty (as described above); and transport uncertainty (which includes uncertainties in instrument wind speed and wind direction, atmospheric stability parameterization, and placement of the sonic anemometer relative to the leak location). These terms are added in quadrature (see Table S1).

The off-diagonal elements in **R** are populated with covariance terms related to measurement proximity in time, with an exponential decay function with a length scale of 10 minutes. This covariance results from the expected covariance of transport uncertainties and expected covariance of measurement uncertainties (due, for example, to expected temporal autocorrelations in turbulence, which affects measurement return power).

| Components of model data mismatch | | 1.1 km beam | 570 m beams |
|---|---|---|---|
| $\sigma_{measurement}$ | (mean value of vector) | 3.0 ppb | 4.22 ppb |
| $\sigma_{transport}$ | (scalar value) | 1.3 ppb | 2.0 ppb |
| $\mathbf{R}_{Inv\text{-}bkg}$ | (mean value of diagonal elements) | 11.2 ppb$^2$ | 21.9 ppb$^2$ |

SI Table 1. Components of the model-data mismatch term, **R**, for the 1.1 km path tests and the 2-leak tests (570 m beams).

**Prior flux estimate, $s_p$**

The prior emissions estimate state vector $s_p$ is populated with emission rates of 0 g min$^{-1}$. This choice serves the important purpose of not biasing results so as not to trigger unnecessary leak detection and repair efforts. This choice represents an important consideration in inversion studies, in which prior assumptions must be carefully considered so as to best fit the goals of and to minimize biased answers to the particular scientific questions being posed. While future applications of the observing system described here might have different aims (for example, to optimize prior bottom-up estimates of facility or city-wide emissions), the goals of the field tests described here are to identify fugitive emissions of methane. In this particular application, very little (if any) prior knowledge of the hourly, daily, diurnal and monthly time variability of total-pad emissions exists. Therefore, it is important that, in the absence of observational constraints (for example if **H** matrices approach zero under shifting winds), that the posterior emission rate revert to zero and not to an otherwise poorly-constrained non-zero emission rate. This choice therefore leads to a low incidence of false positive leak identification. The trade-off for achieving a low incidence of false positive leak identification is that this choice could result in a low bias of recovered emissions. Given the known absence of methane sinks in the timescales and regions in which the tests were performed, we truncate negative fluxes to a maximum limit of 0 g min$^{-1}$.

**Prior flux covariance matrix, Q**

Prior emissions uncertainty is 1.9 g min$^{-1}$, or the maximum allowable emissions rate from a low-bleed pneumatic valve [30]. Flux uncertainties are parameterized with a 3-hour temporal decorrelation length scale describing exponential decay in time between time steps of fluxes. No spatial correlation between potential leak sites is imposed.

The background uncertainty portion of **Q** includes uncertainties in both background sampling and background "construction" (See Table S2). These values are added in quadrature to form the diagonal portion of the part of the **Q** matrix that is devoted to background uncertainty.

| Components of background uncertainty | | 1.1 km beam | 570 m beams |
|---|---|---|---|
| $\sigma_{bkg\_construction}$ | (scalar value) | 1.6 ppb | 1.6 ppb |
| $\sigma_{bkg\_sampling}$ | (mean value of vector) | 2.9 ppb | 4.1 ppb |

SI Table 2. Components of background uncertainty.

Background $CH_4$ covariance is parameterized with a 10-minute temporal decorrelation length scale between time steps.

**Posterior flux covariance matrix, $\hat{Q}$**

Posterior flux uncertainty, $\hat{Q}$, is calculated as [31]:

$$\hat{Q} = Q - QH^T(HQH^T + R)^{-1}HQ. \qquad (2)$$

This allows the calculation of $\hat{Q}$ analytically [31], not by approximation. Reported 1-σ uncertainties are the combination of diagonal and off-diagonal elements related to each spatiotemporal approximation of the flux. In this case, where no averaging is performed through space or time, the 1-σ uncertainties are the square root of the diagonal elements of $\hat{Q}$.

**Data availability**. All data is available from the authors on request.

**References and Notes**

1. Zavala-Araiza, D. *et al.* Super-emitters in natural gas infrastructure are caused by abnormal process conditions. *Nat. Commun.* **8,** ncomms14012 (2017).

2. Brandt, A. R. *et al.* Methane Leaks from North American Natural Gas Systems. *Science* **343,** 733–735 (2014).

3. Schwietzke, S. *et al.* Improved Mechanistic Understanding of Natural Gas Methane Emissions from Spatially Resolved Aircraft Measurements. *Environ. Sci. Technol.* **51,** 7286–7294 (2017).

4. Lauvaux, T. *et al.* High-resolution atmospheric inversion of urban CO2 emissions during the dormant season of the Indianapolis Flux Experiment (INFLUX). *J. Geophys. Res. Atmospheres* **121,** 5213–5236 (2016).

5. Sinclair, L. C. *et al.* Operation of an optically coherent frequency comb outside the metrology lab. *Opt. Express* **22,** 6996–7006 (2014).

6. Truong, G.-W. *et al.* Accurate frequency referencing for fieldable dual-comb spectroscopy. *Opt. Express* **24,** 30495–30504 (2016).

7. Hall, J. L. Nobel Lecture: Defining and measuring optical frequencies. *Rev. Mod. Phys.* **78,** 1279–1295 (2006).

8. Hänsch, T. W. Nobel Lecture: Passion for precision. *Rev. Mod. Phys.* **78,** 1297–1309 (2006).

9. Schliesser, A., Picqué, N. & Hänsch, T. W. Mid-infrared frequency combs. *Nat. Photonics* **6,** 440–449 (2012).

10. Cossel, K. C. *et al.* Gas-phase broadband spectroscopy using active sources: progress, status, and applications [Invited]. *JOSA B* **34,** 104–129 (2017).

11. Coddington, I., Newbury, N. & Swann, W. Dual-comb spectroscopy. *Optica* **3,** 414–426 (2016).

12. Adler, F., Thorpe, M. J., Cossel, K. C. & Ye, J. Cavity-Enhanced Direct Frequency Comb Spectroscopy: Technology and Applications. *Annu. Rev. Anal. Chem.* **3,** 175–205 (2010).

13. Rieker, G. B. *et al.* Frequency-comb-based remote sensing of greenhouse gases over kilometer air paths. *Optica* **1,** 290–298 (2014).

14. Waxman, E. M. *et al.* Intercomparison of Open-Path Trace Gas Measurements with Two Dual Frequency Comb Spectrometers. *Atmos Meas Tech Discuss* **2017,** 1–26 (2017).

15. Tarantola, A. *Inverse Problem Theory Methods for Data Fitting and Model Parameter Estimation*. (Elsevier Sci., 1987).

16. Moore, C. W., Zielinska, B., Pétron, G. & Jackson, R. B. Air Impacts of Increased Natural Gas Acquisition, Processing, and Use: A Critical Review. *Environ. Sci. Technol.* **48,** 8349–8359 (2014).

17. Alvarez, R. A., Pacala, S. W., Winebrake, J. J., Chameides, W. L. & Hamburg, S. P. Greater focus needed on methane leakage from natural gas infrastructure. *Proc. Natl. Acad. Sci.* **109,** 6435–6440 (2012).

18. Pétron, G. *et al.* A new look at methane and nonmethane hydrocarbon emissions from oil and natural gas operations in the Colorado Denver-Julesburg Basin. *J. Geophys. Res. Atmospheres* **119,** 2013JD021272 (2014).

19. Ravikumar, A. P., Wang, J. & Brandt, A. R. Are Optical Gas Imaging Technologies Effective For Methane Leak Detection? *Environ. Sci. Technol.* **51,** 718–724 (2017).

20. Miller, S. M. *et al.* Anthropogenic emissions of methane in the United States. *Proc. Natl. Acad. Sci.* **110,** 20018–20022 (2013).

21. Frankenberg, C., Meirink, J. F., Weele, M. van, Platt, U. & Wagner, T. Assessing Methane Emissions from Global Space-Borne Observations. *Science* **308,** 1010–1014 (2005).

22. Jacob, D. J. *et al.* Satellite observations of atmospheric methane and their value for quantifying methane emissions. *Atmos Chem Phys* **16,** 14371–14396 (2016).

23. Cambaliza, M. O. L. *et al.* Assessment of uncertainties of an aircraft-based mass balance approach for quantifying urban greenhouse gas emissions. *Atmos Chem Phys* **14,** 9029–9050 (2014).

24. Conley, S. *et al.* Methane emissions from the 2015 Aliso Canyon blowout in Los Angeles, CA. *Science* **351,** 1317–1320 (2016).

25. Brantley, H. L., Thoma, E. D., Squier, W. C., Guven, B. B. & Lyon, D. Assessment of Methane Emissions from Oil and Gas Production Pads using Mobile Measurements. *Environ. Sci. Technol.* **48,** 14508–14515 (2014).

26. Yacovitch, T. I. *et al.* Mobile Laboratory Observations of Methane Emissions in the Barnett Shale Region. *Environ. Sci. Technol.* **49,** 7889–7895 (2015).

27. Pétron, G. *et al.* Hydrocarbon emissions characterization in the Colorado Front Range: A pilot study. *J. Geophys. Res. Atmospheres* **117,** D04304 (2012).

28. Roscioli, J. R. *et al.* Measurements of methane emissions from natural gas gathering facilities and processing plants: measurement methods. *Atmos Meas Tech* **8,** 2017–2035 (2015).

29. Turner, D. B. A Diffusion Model for an Urban Area. *J. Appl. Meteorol.* **3,** 83–91 (1964).

30. US EPA, O. Inventory of U.S. Greenhouse Gas Emissions and Sinks: 1990-2015. *US EPA* (2016). Available at: https://www.epa.gov/ghgemissions/inventory-us-greenhouse-gas-emissions-and-sinks-1990-2015. (Accessed: 25th August 2017)

31. Yadav, V. & Michalak, A. M. Improving computational efficiency in large linear inverse problems: an example from carbon dioxide flux estimation. *Geosci. Model Dev.* **6,** 583–590 (2013).

**Acknowledgements:** We acknowledge financial support from the Advanced Research Projects Agency-Energy, Department of Energy (award DE-AR0000539), the Department of Energy (award DE-FE0029168), the Defense Advanced Research Projects Agency under the DSO SCOUT program, and the National Institute of Standards and Technology.

**Author contributions:** G.B.R, I.C., C.S. and N.R.N. organized project. S.C., R.W., K.C., E.B., and G.W.T. set up field experiment. S.C. and R.W. conducted measurements. S.C. and C.B.A. analyzed data. F.G. provided guidance on measurement analysis. K.P. provided guidance on inversions. All authors provided comments and revisions on the paper. S.C., C.B.A., and G.B.R wrote the paper.

**Competing financial interests:** The authors declare no competing financial interests.

**Figure 1**. Regional source monitoring with a centralized dual-comb spectrometer **(a)** measures trace gas absorption over an array of long-distance beam paths. **(b)** Time-resolved trace gas concentrations are determined from fits to the absorption spectra with ppb-km sensitivity and stability. **(c)** An atmospheric transport model and inversion determines source location and time-resolved emission rate.

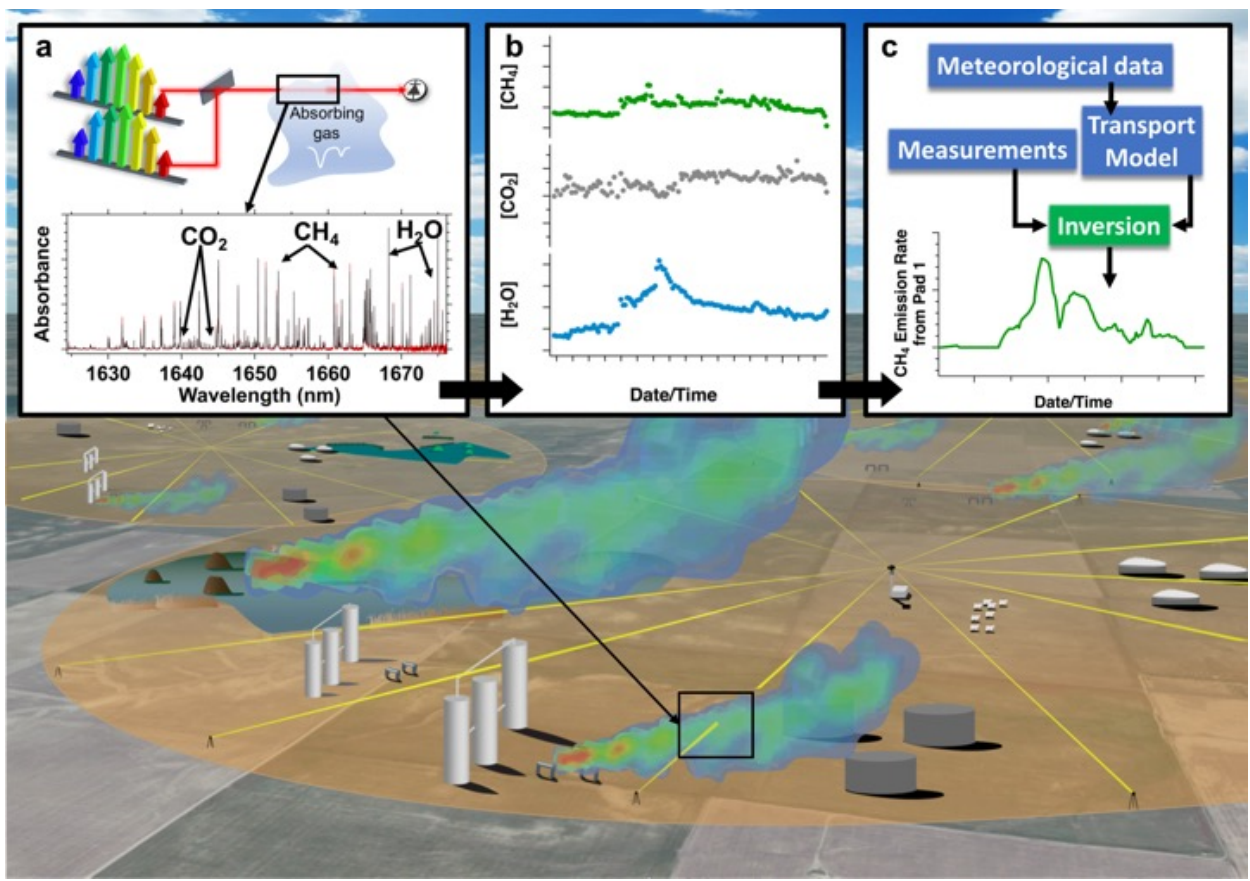

**Figure 2**. Overview of the field site. (**a**) Table Mountain field site location. (**b**) Zoomed view of the site including mobile laboratory (yellow square) and the area over which tests were conducted (black circle). (**c**) Field deployed DCS, (**d**) gimbal/telescope, and (**e**) retroreflector.

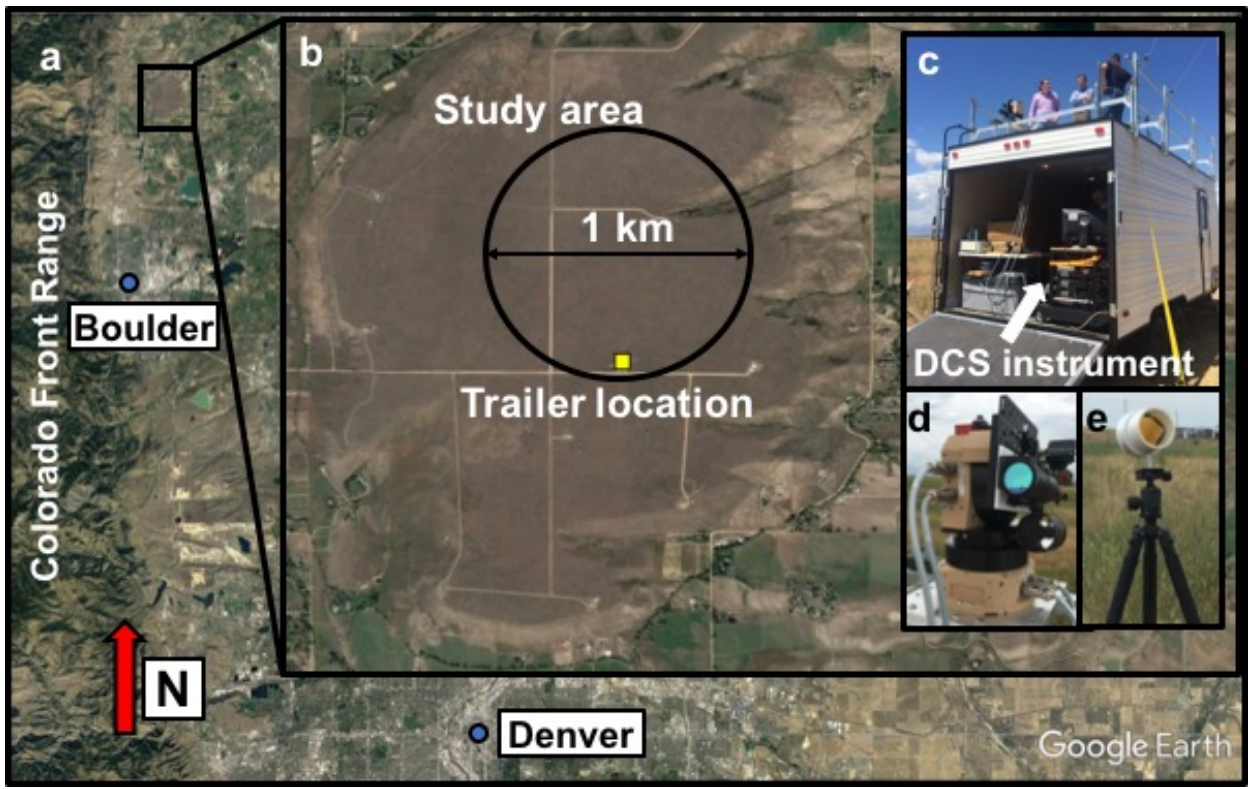

**Figure 3**. Detection of a small, time-varying methane source from 1 km. **(a)** Map showing the site configuration including retroreflectors (blue diamonds) and source (red circle). **(b)** Methane concentrations measured on beam paths shown in **(a)**. The light blue line denotes the background measurement (the upwind beam depends on wind direction). **(c)** Retrieved emission rate (blue line, error bars are 1-σ posterior uncertainty), compared with true emission rate (black dotted line). Also shown is the prior estimate of the emission (thin black line) used in the inversion and the average values for both the true emission rate (maroon dashed line) and the posterior (grey solid line with mean uncertainty).

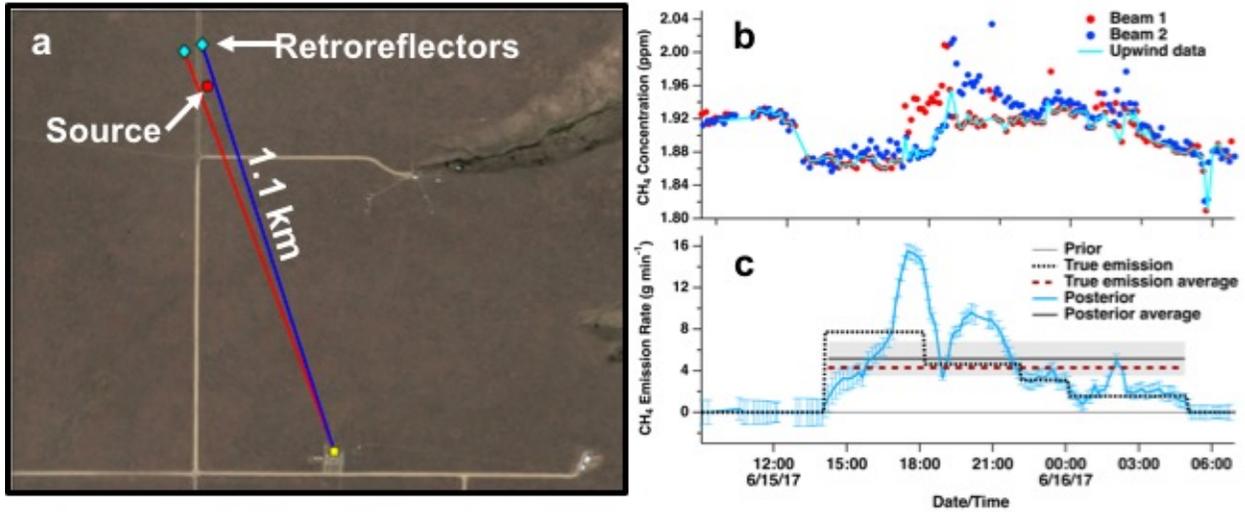

**Figure 4**. Detection of two sources from among multiple potential sources. Layout of **(a)** and **(b)** in this figure follow that of Fig. 3. **(c)** True emission rates (sources 2 and 4: solid grey lines; sources 1, 3, and 5: dotted black lines) and retrieved emission rates (sources 1: grey squares, 2: red diamonds, 3: orange diamonds, 4: purple hourglasses, 5: gold asterisks).

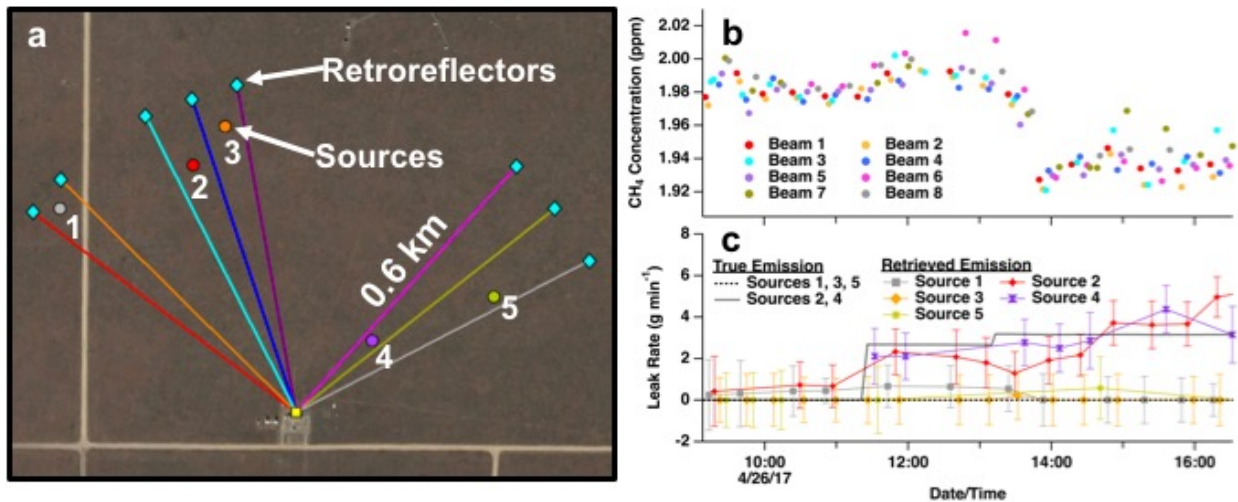

**Supplementary Information**

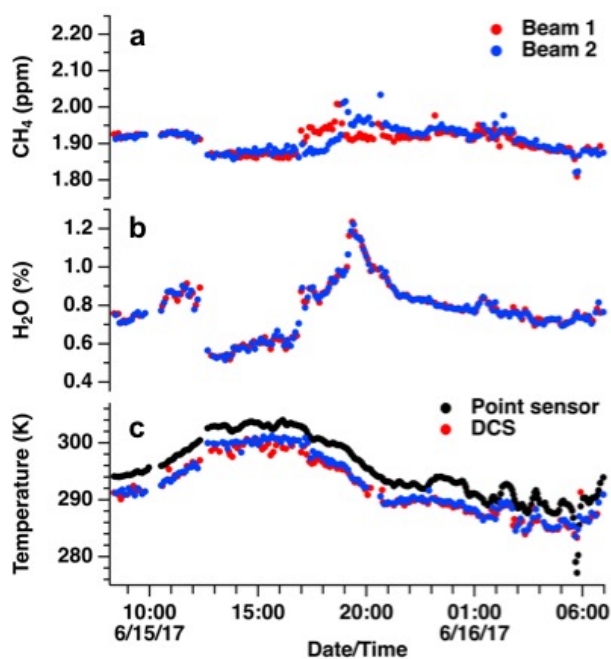

**Figure S1:** Concentration measurements from the long path experiment: **(a)** $CH_4$, **(b)** water vapor; **(c)** temperature. The color coding of the traces distinguishes the measurements by retroreflector, i.e. which beam path the measurement represents.

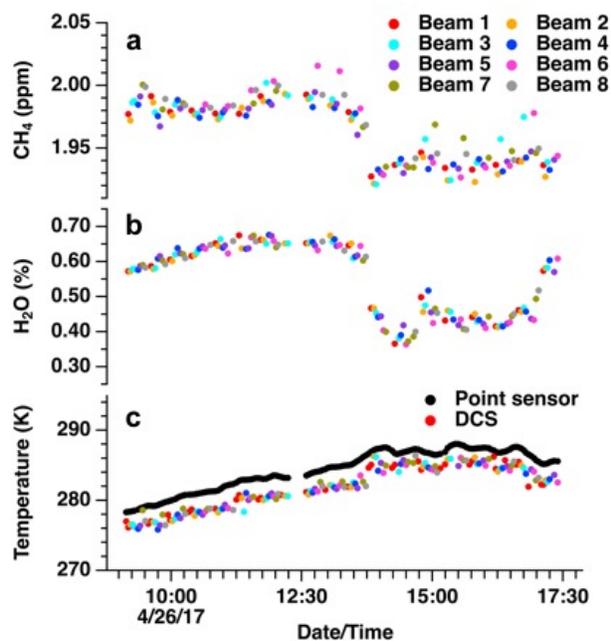

**Figure S2:** Concentration measurements from the multi-source experiment: **(a)** $CH_4$, **(b)** water vapor; **(c)** temperature. The color coding of the traces distinguishes which retroreflector was targeted, i.e. which beam path the measurement represents.

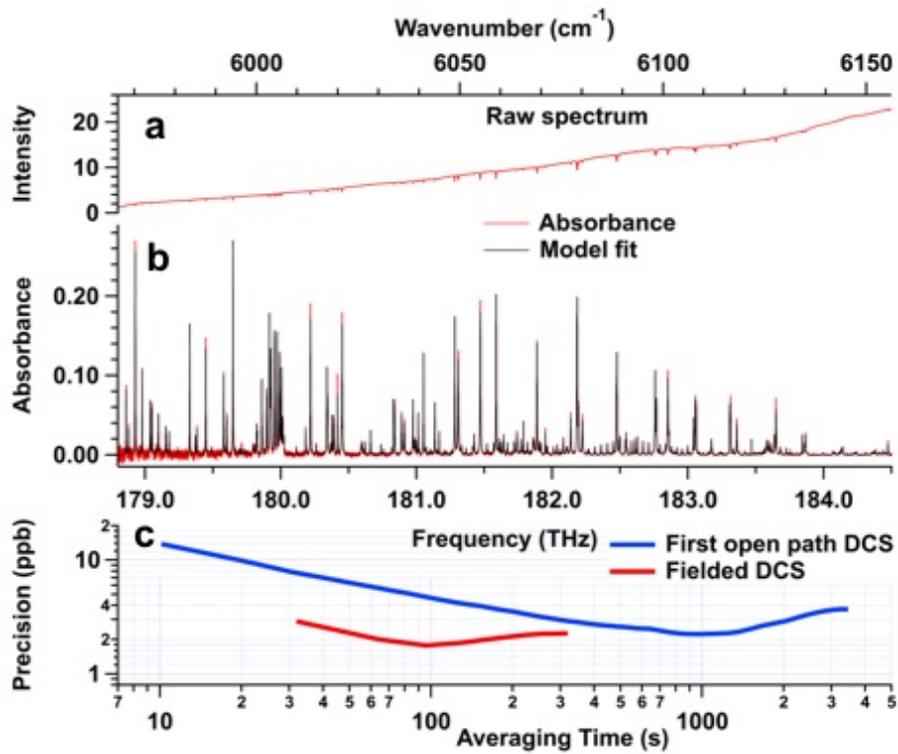

**Figure S3**. **(a)** Raw spectrum, **(b)** fit to absorbance model, and **(c)** allan deviation for background data collected during the long path measurement experiment. Also included in **(c)** is an allan deviation trace from the first open path DCS measurements [13].

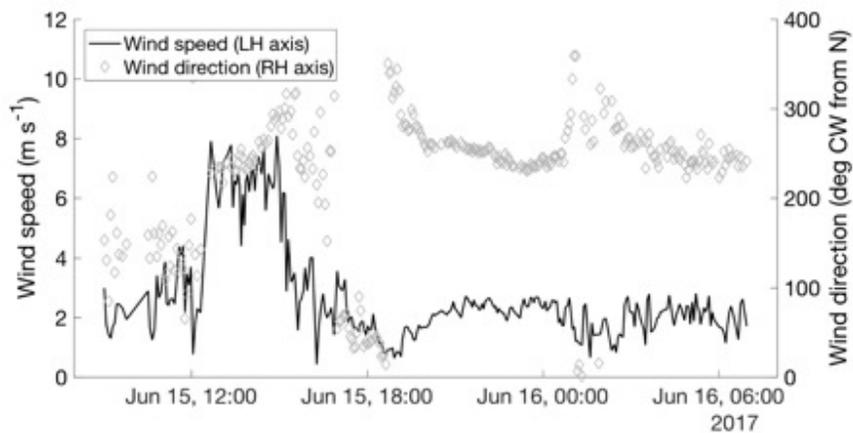

**Fig. S4.** Meteorological data from 1 km beam path controlled leak tests.

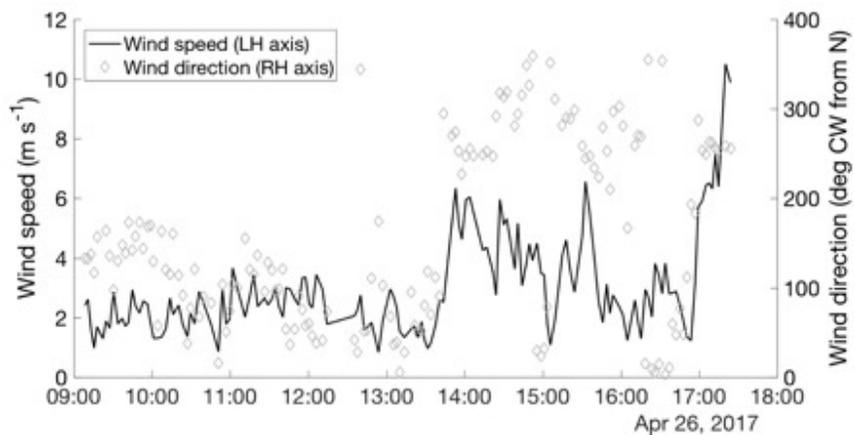

**Fig. S5.** Meteorological data from multi-source controlled leak tests